\documentclass[aps,prl,twocolumn,superscriptaddress]{revtex4-1}
\usepackage{graphicx}
\usepackage{subfig}
\usepackage{floatrow}
\usepackage{amsmath}
\usepackage{hyperref}
\usepackage{floatrow}
\usepackage{siunitx}
\usepackage {ulem}

\usepackage{ragged2e}
\usepackage{dcolumn}
\usepackage{bm}

\begin{document}

\preprint{APS/123-QED}

\title{Breakdown of the quantum anomalous Hall effect under microwave drives}

%

\author{Torsten Röper}
\author{Daniel Rosenbach}
\affiliation{II. Physikalisches Institut, Universit\"at zu K\"oln, Z\"ulpicher Str. 77, D-50937 K\"oln, Germany}
\author{Achim Rosch}
\affiliation{Institute of Theoretical Physics, Universit\"at zu K\"oln, Z\"ulpicher Str. 77, D-50937 K\"oln, Germany}
\author{Alexey A. Taskin}
\author{Yoichi Ando}
\author{Erwann Bocquillon}%
 \email{bocquillon@ph2.uni-koeln.de}
 \affiliation{II. Physikalisches Institut, Universit\"at zu K\"oln, Z\"ulpicher Str. 77, D-50937 K\"oln, Germany}

\date{\today}

\begin{abstract}
Quantum anomalous Hall (QAH) insulators exhibit chiral dissipationless edge states without an external magnetic field, making them a promising material for quantum metrology and microwave applications. However, the breakdown of the zero-resistance state at low currents hinders progress. We investigate and characterize this breakdown under microwave fields (1–25 GHz) by measuring the increase of longitudinal resistance in RF Hall bars and RF Corbino devices made from V-doped (Bi,Sb)$_2$Te$_3$ films. Our results point to the role of heating of electron-hole puddles under microwave irradiation, thereby fostering hopping transport. Our work offers insights critical for GHz-range QAH applications.

\end{abstract}

\maketitle


\paragraph{\label{sec:Intro} Introduction --} 
    The quantum anomalous Hall (QAH) effect is characterized by a single chiral topological edge state, yielding a vanishing longitudinal resistance and a quantized Hall resistance without external magnetic field \cite{chang_experimental_2013,chang_high-precision_2015,Yu2010}. Materials that inherit the QAH effect have significant potential for applications in quantum metrology \cite{Bestwick-2015,gotz2018,Fox-2018,fijalkowski2024} and for non-reciprocal microwave components \cite{viola_hall_2014,mahoney_zero-field_2017}. Additionally, coupling QAH systems with superconductors paves an alternative route to topological quantum computation \cite{uday2023,beenakker_deterministic_2019,beenakker_electrical_2019}.
    However, a key limitation in using QAH insulators is the breakdown of the zero-resistance state at low temperatures or under small biases, marking the onset of dissipative bulk transport \cite{ lippertz-2022,fijalkowski2021a,fijalkowski2024}. This breakdown is attributed to various mechanisms: variable-range hopping \cite{Kawamura-2017}, bootstrap electron heating \cite{Fox-2018} and percolation through charge puddles (locally conducting regions which form in the bulk due to Coulomb disorder \cite{Skinner2013,chang_experimental_2013,bagchi2019,Brede2024}) driven by electric fields \cite{lippertz-2022}. For example, low- and high-frequency longitudinal dissipation exhibit signatures of variable-range hopping (VRH) \cite{chang-2015,lippertz-2022,Kawamura-2017,Fox-2018}, governed by hopping between localized states assisted by thermally excited phonons, and has been observed in various disordered narrow gap semiconductors\cite{Ambegaokar,efros_1975,baranovski_charge_2006,Shklovskii2024}.

   The breakdown of QAH systems driven by microwave signals remains however mostly unexplored. While disconnected, isolated puddles do not conduct in the DC limit, capacitive coupling between them allow for finite transport at higher frequencies. Much below the breakdown, the role of puddles on the physics of edge states is in fact now well understood, thanks to recent narrow and broadband microwave studies at GHz frequencies relevant for potential applications \cite{mahoney_zero-field_2017,Dartiailh2020,Kamata2022,roeper2024}. However, breakdown at such frequencies is much less explored. Two main scenarios can be envisioned. On the one hand, RF photons can promote hopping between sites and trigger photon-assisted variable-range hopping \cite{Keiper1978,zvyagin_hopping_1978}. On the other hand, RF signals can generate Joule heating, thereby elevating the temperature and indirectly increasing the DC bulk transport. This heating mechanism is all the more relevant as the conductivity through the puddle network, and therefore the dissipation, is enhanced for GHz-frequencies \cite{roeper2024}.
    
      \begin{figure}[t]
        \centering
        \includegraphics[width=\textwidth]{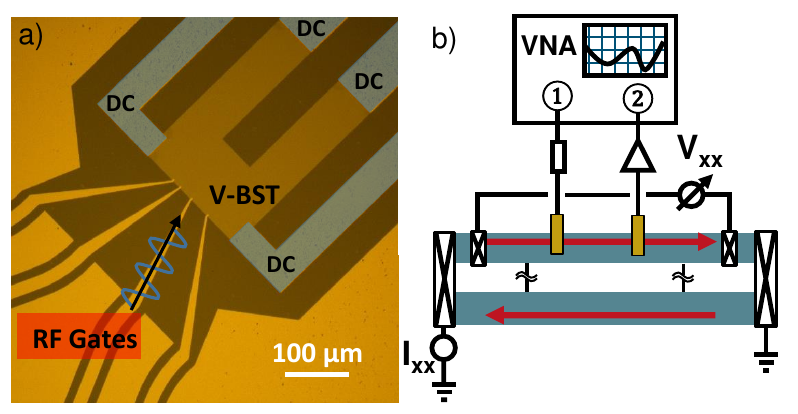}
        \caption{\justifying\textbf{Experimental setup and DC characterization :} a) Microscope image of Sample A in a RF Hall bar geometry. The MESA is false-colored in dark yellow, the ohmic contacts in blue, the RF finger gates are shown in the bottom left. All remaining contacts are ohmic contacts. b) Schematic of the microwave measurement setup. A vector network analyzer (VNA) is used as a signal generator and connected via broadband coaxial cables and a coplanar waveguide. Longitudinal resistance $R_{xx}$ is measured using a four-terminal lock-in configuration.}
        \label{fig:device_setup}
    \end{figure}
      

  \begin{figure*}[htp!]
        \centering
        \includegraphics[width=\textwidth]{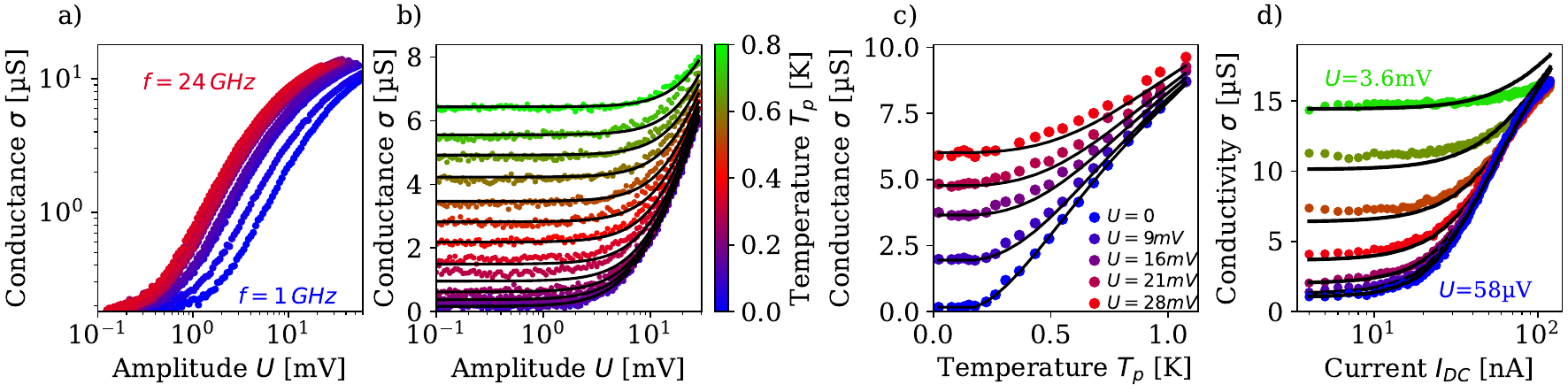}
        \caption{\justifying\textbf{Breakdown driven by RF excitation and temperature:} a) Conductance $\sigma$ as a function of the RF amplitude $U$. The measurement was performed at $B=\SI{1}{\tesla}$, $T_p\simeq\SI{20}{\milli\kelvin}$ on Sample A. b) Conductance $\sigma$ as a function of the RF amplitude $U$ for different temperatures from \SI{20}{\milli\kelvin} up to \SI{800}{\milli\kelvin}, at $f=\SI{8}{\giga\hertz}$. The black lines correspond to the fit of the Joule heating model yielding $\alpha=\num{3.5(0.1)}$, $\Sigma=\SI{32.6(7)}{\micro\watt\per\kelvin^\alpha}$ and a phonon temperature $T_p$, which is set to the temperature of the sample holder (taken on Sample A). c) Conductance $\sigma$ as a function of temperature $T_p$ for RF amplitudes $U$ between 0 (blue) and \SI{44}{\micro\volt} (red), at $f=\SI{8}{\giga\hertz}$. The blue curve is fitted with a model for variable-range-hopping transport yielding an activation temperature of $T_0=\SI{17(0.3)}{\kelvin}$ (taken on Sample A). d) Onset of conductivity $\sigma$ as a function of the DC current $I_{\rm DC}$ while applying a microwave drive at \SI{8}{\giga\hertz} and varying amplitude ($U$=0.1,0.2,0.5,0.9,1.8,\SI{3.6}{\milli\volt}). The measurements where taken on Sample B at $T\approx\SI{20}{\milli\kelvin}$ and $B=\SI{2}{\tesla}$. The black lines correspond to a fit with the Joule heating model, using the fit parameters obtained in DC ($\Sigma=\SI{37(2)}{\nano\watt\per\kelvin^\alpha}$, $\alpha=\num{4.5\pm0.1}$) as fixed parameter and fitting the impedance $Z$ as a free fit parameter, yielding $Z=\SI{3280(60)}{\ohm}$. }
        \label{fig:breakdown}
\end{figure*}
    
    In this study, we investigate the breakdown of the QAH state in the presence of microwave signals and assess the role of hopping. To address this, we fabricate samples comprising vanadium-doped (Bi,Sb)$_2$Te$_3$ (V-BST) Hall bar and Corbino devices and coplanar waveguides to apply microwave signals. To realize a frequency-independent dissipation probe and study different frequencies on equal footing, we use the longitudinal conductance $\sigma=\sigma_{xx}$ measured in DC as a probe, while applying microwave fields between 1 and \SI{25}{\giga\hertz}. We find that the electric field necessary for a breakdown strongly decreases with its frequency. Analyzing the dependence of $\sigma$ on RF amplitude, frequency, and temperature, we assess different models and finally ascribe the phenomenon to RF-induced heating and a subsequent increase of the electron and lattice temperature. To validate our findings, we present a microscopic model of the hopping transport that reasonably describes the RF amplitude and frequency dependence of the edge state resistance.

\paragraph{\label{sec:Device} Experimental Setup --}

    \begin{table}[b!]
    \caption{\label{tab:Device-list} \textbf{Overview of samples}: Hall resistance $R_{yx}$, longitudinal resistance $R_{xx}$, and the geometry of all devices}
    \begin{ruledtabular}
    \begin{tabular}{cccc}
    \textrm{Sample}&
    \textrm{$R_{yx}/R_{K}$} & $R_{xx}$ [\si{\ohm}] & Geometry\\
    \colrule
    A (Feb6a) &  $\mathrm{1.00 \pm 0.01}$ & $\mathrm{100 \pm 10}$ & Hall bar\\
    B (Jul5a) &  $\mathrm{1.00 \pm 0.02}$ & $\mathrm{90 \pm 30}$ & Corbino\\
    C (Jul31a) &  $\mathrm{1.00 \pm 0.02}$ & $\mathrm{120 \pm 60}$ & Hall bar\\
    \end{tabular}
    \end{ruledtabular}
    \end{table}

    We employed two distinct device designs to investigate edge-state dissipation under microwave fields: RF Hall bars and RF Corbino rings. Both designs feature ohmic contacts for DC characterization and narrow capacitive contacts for coupling microwave signals. The RF Hall bar geometry (Samples A and C) matches the one described in our previous work \cite{roeper2024}. A microscope image of Sample A is shown in Fig.~\ref{fig:device_setup}a. Sample B in the RF Corbino geometry is based on the conventional Corbino annular geometry, and provides a direct measure of the bulk conductance, isolating edge contributions (see \cite{supplement}).
    All devices were fabricated using \SI{8}{\nano\meter} thick films of V-BST grown via molecular beam epitaxy, capped with a \SI{4}{\nano\meter} layer of $\rm Al_2O_3$. The fabrication process is identical to that described in \cite{roeper2024}. Table~\ref{tab:Device-list} provides an overview of the devices. Each device demonstrated a quantized Hall resistance ($R_{yx}=h/e^2$) with an accuracy of \SI{2}{\percent} and longitudinal sheet resistance values ranging between \SI{90}{\ohm} and \SI{120}{\ohm} in the low voltage regime.\\
    The measurements are performed in a dilution refrigerator with a base temperature below \SI{20}{\milli\kelvin}, in the configuration shown in Fig.\ref{fig:device_setup}b. To isolate the frequency dependence of the breakdown of the edge state transport, we are measuring the longitudinal resistance in DC to realize a frequency-independent probe of the breakdown. Therefore, we source a small probe current (\SI{1}{\nano\ampere}) well below the breakdown current and determine the longitudinal resistance $R_{xx}$ from the voltage drop measured between two other ohmic contacts located between the source and drain contacts. We calculate the longitudinal bulk conductance $\sigma$ from the longitudinal resistance $R_{xx}$ and the transverse (Hall) resistance $R_{yx}$:
    \begin{equation}
        \sigma = \sigma_{xx} = \frac{R_{xx}}{R_{xx}^2+R_{yx}^2}
    \end{equation}
   In the Corbino geometry, we measure the bulk current to confirm that the increase in resistance is due to bulk conductance, as detailed in the Supplementary material \cite{supplement}.
    Microwave signals are applied via the narrow capacitive contacts. These signals originate from a vector network analyzer (VNA) at room temperature, connected to the device through coaxial cables.
    The microwave power $P_{\rm in}$ at the sample (or equivalently the amplitude $U$) is first calibrated by measuring with the VNA the transmission through a reference sample with a \SI{50}{\ohm}-matched coplanar waveguide. This calibration was further validated using a tunnel diode detector, as detailed in \cite{supplement}. In the rest of the study, we use the VNA solely as a signal generator.

    \begin{figure}[t]
        \centering
        \sidesubfloat[\centering]{{\includegraphics[height=5.5cm,keepaspectratio=true]{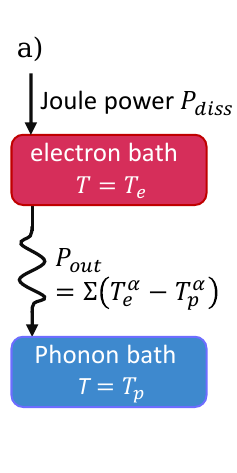} }}
        \sidesubfloat[\centering]{{\includegraphics[height=5.5cm,keepaspectratio=true]{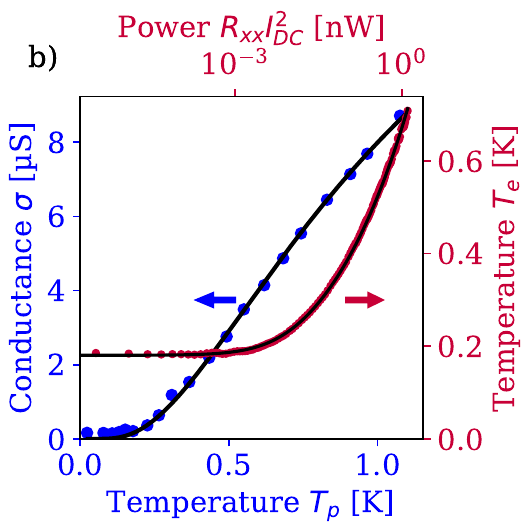} }}
        \caption{\justifying\textbf{Joule heating and electronic thermometry:} a) Joule heating model: The current dissipated heat $P_{\rm in}$, that increases the electronic temperature $T_e$. The electrons are cooled via phonon emission with the power $P_{\rm out}$. b) The blue dotted curve shows the bulk conductance $\sigma$ (left axis) with the refrigerator temperature $T_p$ (bottom axis). The curve is used to calibrate the electron temperature $T_e$ using the VRH law. Fitting yields $T_0=\SI{17(0.3)}{\kelvin}$ and $\sigma_0=\SI{490(20)}{\micro\siemens\kelvin}$. The red dotted curve shows the calculated electronic temperature $T_e$ (right axis) as a function of the dissipated power $P_{\rm in}=R_{xx} \,I_{\rm DC}^2$ (top axis) under a DC current is shown. The black solid line shows a fit to Eq.(\ref{eq:Te}) with $\alpha=4$, yielding $\Sigma=\SI{4.94(0.02)}{\nano\watt\per\kelvin^4}$.}
       \label{fig:temp}
    \end{figure}

\paragraph{\label{sec:Frequency} Experimental observations --}
    In the main text, we focus on the results obtained on Sample A and Sample B, more results are shown in \cite{supplement}. First, we discuss the measurement of the conductance $\sigma$ as a function of the amplitude $U$ of the microwave signal, shown in Fig.~\ref{fig:breakdown}a for frequencies ranging from \SI{1}{\giga\hertz} (blue dots) up to \SI{25}{\giga\hertz} (red dots) with logarithmic axes. All curves exhibit a similar residual conductance $\sigma$ at low RF amplitudes $U$, and then a sharp increase with a threshold RF amplitude $U_{\rm BD}$ that shifts to lower amplitudes with increasing frequencies. \\

In addition to varying the frequency $f$, we also measured the conductance $\sigma$ as a function of the RF amplitude $U$ for various sample temperatures $T_p$ (Fig.\ref{fig:breakdown}b). We observe at low $U$ a constant value of $\sigma$, which shifts to higher values. At high power, $\sigma$ increases with $U$ signaling the breakdown induced by the RF drive. However, its effects are less visible at high temperatures, and pushed to increasingly higher values of $U$.
Additionally, we show in Fig. \ref{fig:breakdown}c the measurement of $\sigma$ as function of sample temperature $T_p$ for various values of $U$. We observe at low $T_p$ a constant value of $\sigma$ that increases with increasing $U$. At higher temperatures, $T_p$, $\sigma$ shows an increase with $U$ at a threshold value of $T_p$ which itself increases with increasing $U$. At very high $T_p$ we find that all curves overlay. Finally, we probe the breakdown of the QAH state under mixed large DC and RF biases, by measuring the conductance $\sigma$ as a function of the current $I_{\rm DC}$ for various values of the RF amplitude $U$. At low $I_{\rm DC}$, the conductance saturates at a minimum value which depends on $U$. For larger $I_{\rm DC}$, it strongly increases, signaling the breakdown. As before, all curves overlay independently of $U$ for very large $I_{\rm DC}$ deep in the breakdown regime.

These four types of measurements differently illustrate the action of RF signals on the bulk conductance of our samples, ultimately leading to the breakdown of the QAH regime. They constitute the main experimental findings of this article. We develop in the next sections a simple model based on Joule heating to explain these observations. We here note that we have, in the course of the project, explored other models of variable-range hopping. We show in particular in the Supplementary material \cite{supplement} that photon-assisted variable-range hopping \cite{zvyagin_hopping_1978,Keiper1978} or models developed for quantum Hall transitions \cite{Bennaceur2012, Polyakov1993} do not provide a satisfactory description of our observations.



\paragraph{\label{sec:theory} Model for Joule-induced heating --}
  
In this section, we show that the results are well reproduced by a simple model accounting for the heating of the electron bath by the RF excitation, which in turn triggers variable-range hopping. The model is depicted in Fig.\ref{fig:temp}a. The electron bath, at a uniform temperature $T_e$ is coupled to the lattice phonons at temperature $T_p$ via electron-phonon coupling. At equilibrium, the electron temperature $T_e$ is set by the balance of the power $P_{\rm out}$ taken away through the lattice by phonons, and the power $P_{\rm diss}$ dissipated in the electron bath via the Joule heating, either from the applied DC current or by absorbing the GHz radiation. At this stage, the model does not specify the absorption mechanism, but the prime candidate is absorption from electron-hole puddles. The model also entails approximations on the uniformity of the temperatures $T_e$ and $T_p$, or on the geometry of the RF field. We discuss these approximations and hypotheses in \cite{supplement}.

We first study the change of $\sigma$ as function of the temperature of the refrigerator (assumed to be equal to $T_p$), as shown in Fig.\ref{fig:temp}b. In these conditions, assuming thermal equilibrium between the electron and phonon baths ($T_e=T_p$), the behavior of $\sigma$ is well described by phonon-assisted variable-range hopping (VRH) transport and follows the expression $\sigma(T_e)=\frac{\sigma_0}{T_e}\exp\left(-(T_0/T_e)^{1/2}\right)$ \cite{roeper2024, martinez_edge_2023}, which produces an activation temperature $T_0=\SI{17(0.3)}{\kelvin}$. In a regime where electrons equilibrate, we can use this relation as an effective thermometer sensitive to the electron temperature $T_e$.
In semiconducting devices at low temperatures, cooling through electron-phonon scattering is known to follow the typical power law $P_{\rm out}=\Sigma\big(T_e^\alpha-T_p^\alpha)$ with $\Sigma$ a constant, and the exponent $\alpha$ in the range $3$ to $5$ \cite{jezouin2013, lebreton2022, Ferguson2025}. Equating $P_{\rm diss}=P_{\rm out}$ at thermal equilibrium yields:
\begin{equation}
    T_e=\Big(T_p^\alpha+\frac{P_{\rm diss}}{\Sigma}\Big)^{1/\alpha}\label{eq:Te}
\end{equation}

In the DC regime, as represented in Fig.\ref{fig:temp}b, we can verify that the application of Joule power $P_{\rm diss}=R_{xx}\,I_{\rm DC}^2$ yields an excellent agreement with Eq.(\ref{eq:Te}), setting $\alpha=\num{4}$, and for $\Sigma=\SI{4.94(0.02)}{\nano\watt\per\kelvin^4}$. We show in \cite{supplement}, however, that slightly smaller exponents are obtained for the other samples, possibly due to a different thermal coupling to the substrate. We note that the residual conductance at $T_p\to0$, not taken into account by VRH, results in an irrelevant temperature offset $T_e\simeq\SI{190}{\milli\kelvin}$.

\begin{figure}[ht!]
    \centering
    \includegraphics[width=\textwidth]{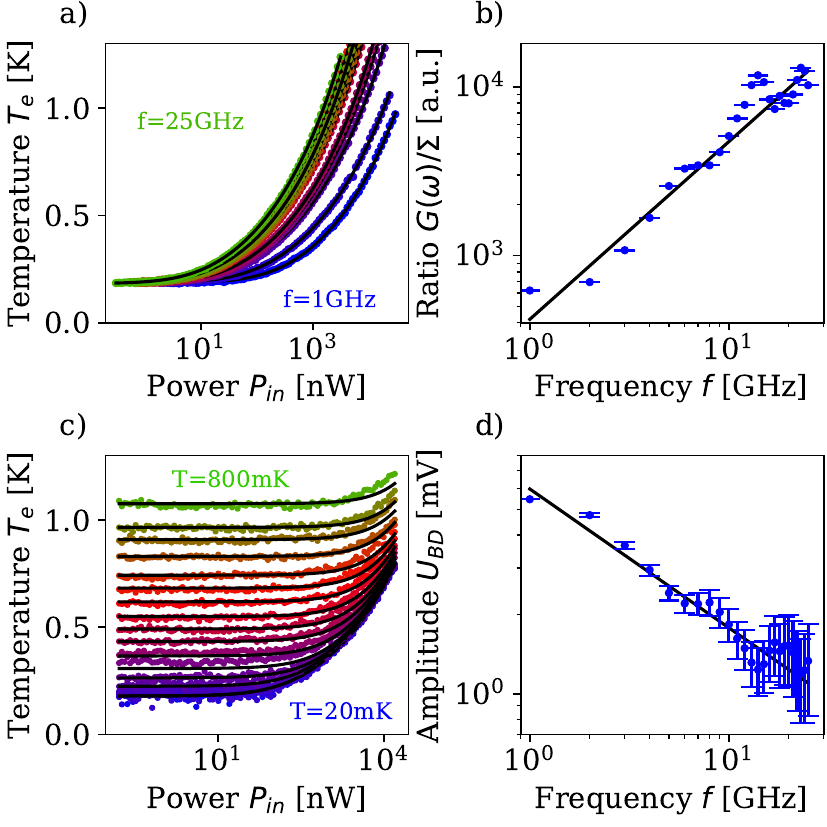}
    \caption{\justifying\textbf{RF-induced heating:} a) Calculated electronic temperature $T_e$ as a function of the applied microwave power $P_{\rm in}$ for frequencies between 1 and \SI{25}{\giga\hertz}. The black lines show the fitted curve of the Joule heating model, yielding $\alpha=\num{3.5\pm0.1}$ and the ratio $G(\omega)/\Sigma$ shown in Fig.~4b. b) Frequency dependence and fit of the parameter $G(\omega)/\Sigma=A \omega^s$, yielding $A=\SI{420(50)}{\kelvin^\alpha\siemens\per\giga\hertz^s \per\watt}$ and $s=\num{1.05\pm0.05}$. c) The electronic temperature $T_e$ obtained via the conductance $\sigma$ is plotted as a function of the applied microwave power $P_{\rm in}$ for various phonon temperatures $T_p$, at $f=\SI{8}{\giga\hertz}$. The black lines correspond to simulations using Eq.(2) and the parameters obtained from Fig.4a. d) The breakdown RF amplitude $U_{\rm BD}$ (defined at $\sigma=\sigma_{\rm BD}=\SI{1}{\micro\siemens}$) as a function of the microwave frequency $f$. The black line corresponds to the predicted behavior $U_{\rm BD}\propto 1/\sqrt{\omega}$ of the Joule heating model, extracted from the fit shown in b).}
    \label{fig_4}
\end{figure}

\paragraph{\label{sec:theory2} Heating from microwaves --}
In the RF regime, the power $P_{\rm in}$ applied on the gate is partially absorbed by the sample. We write the dissipated power as $P_{\rm diss}(\omega)=G(\omega) U(\omega)^2$ where $G(\omega)$ is the effective conductance dissipating the amplitude $U(\omega)$ applied to the gate at angular frequency $\omega=2\pi/ f$. We find again excellent agreement between $T_e$ and $P_{\rm in}(\omega)$ as shown in Fig.\ref{fig_4}a, where we used the ratio $G(\omega)/\Sigma$ as a fitting parameter. The exponent $\alpha$ shows no clear frequency dependence (fit parameters between $\num{3.2}$ and $\num{3.5}$), but is slightly smaller than the one found in the DC regime (in all measured samples, see \cite{supplement}  for details). Under the assumption that $\Sigma$ is approximately constant in this frequency range much below the Debye temperature, we observe (Fig.\ref{fig_4}b) that $G(\omega)$ can be well fitted by $G(\omega)/\Sigma\propto\omega^s$, with an exponent $s=\num{1.05\pm0.05}$. Thus, absorption increases with increasing frequencies. This is consistent with a scenario of absorption by finite-size puddles for frequencies smaller than the inverse Thouless time, $f\lesssim 1/\tau_{\rm Th}$. This timescale $\tau_{\rm Th}=DL_p^2$ is set by the size of $L_p$ of the puddles and the local diffusion constant $D$ \cite{bagchi2019}. This relation agrees with experimental reports on other strongly disordered systems and confirms the role of disorder in our samples \cite{baranovski_charge_2006,hill1979}.

The analysis of the behavior of $\sigma$ as a function of $U$ as described above allows the extraction of the model parameters (namely $\alpha$ and  $G(\omega)/\Sigma$) for the base temperature $T_p\simeq T_e\simeq\SI{20}{\milli\kelvin}$. In turn, we can then predict the behavior of $T_e$ and $\sigma$ when the sample temperature $T_p$ is varied. We obtain excellent agreement without any further fitting parameters, as shown in Figs. \ref{fig:breakdown}b, \ref{fig:breakdown}c and \ref{fig_4}c. This contrasts in particular with the other studied models, in which the variation of $T_p$ led to strong discrepancies between model and data (see \cite{supplement}), and strengthens our interpretation based on RF-induced heating. To simulate the data taken under mixed DC and RF biases, we set the exponent $\alpha=4$ and write $P_{\rm diss} = R_{xx}I_{\rm DC}^2+ U^2/Z$ the total dissipated power, with $Z$ a free parameter (dimensioned as an impedance). Though the agreement is not as good (see Fig.\ref{fig:breakdown}d), the trends are adequately reproduced. This also suggests that heating plays a role in the breakdown under DC biases, as pointed out in previous works \cite{Rosen_2022}.

Finally, we come back to the initial purpose of the article, namely the characterization of the QAH breakdown thresholds when driven by RF signals. According to our model, the breakdown threshold conductance $\sigma_{\rm BD}$ is reached for a constant heating power, thus defining a breakdown threshold amplitude $U_{\rm BD}\propto 1/\sqrt{\omega}$. To verify this simple prediction, we define in Fig.\ref{fig:breakdown} the breakdown threshold amplitude $U_{\rm BD}$ as the amplitude $U$ at which $\sigma=\sigma_{\rm BD} = \SI{1}{\micro\siemens}$. This threshold is chosen to lie in an intermediate regime of the breakdown (approx. 5 times the residual bulk conductance). Fig.~\ref{fig_4}d shows a good agreement between the extracted breakdown RF amplitude $U_{\rm BD}$ and our model prediction (based on the extracted fit parameters). This prediction is a valuable asset in the pursuit of lossless high-frequency devices based on the QAH effect \cite{viola_hall_2014,mahoney_zero-field_2017}.

\paragraph{\label{sec:discussion} Summary and conclusions --}
This study provides a comprehensive analysis of the breakdown of the quantum anomalous Hall effect at high frequencies. Combining electronic transport via variable-range hopping and thermal relaxation due to electron-phonon coupling, we successfully model our data. Previous works on the breakdown of the QAH state sketched different scenarios, some favoring the role of puddles and others the effects of heating. Our results actually indicate that both aspects combine and that heating occurs at high frequency in charge puddles forming in V-BST QAH insulators. It mirrors recently reported findings in QAH materials \cite{Ferguson2025} but more clearly reveals the role of puddles and AC dissipation. The breakdown is also significantly different from that of quantum Hall systems, where slow relaxation dynamics \cite{sagol_time_2002,kalugin_relaxation_2003,akera_slow_2009} and Landau level physics are involved \cite{eaves1986,yang2018}. Finally, our model clearly predicts breakdown thresholds for the QAH state at high frequencies, which decrease with frequency as $1/\sqrt{\omega}$. Together with residual dissipation in the low amplitude regime \cite{roeper2024}, this puts further stringent conditions on the viability of the QAH effect to realize lossless high-frequency quantum devices \cite{viola_hall_2014,mahoney_zero-field_2017}.\\

The supporting data and codes for this article are available from Zenodo \cite{zenodo_data}.

\section{Acknowledgements}
    \begin{acknowledgments}
        We warmly thank B. Shklovskii for his insights. This work has been supported by Germany’s Excellence Strategy (Cluster of Excellence Matter and Light for Quantum Computing ML4Q, EXC 2004/1 - 390534769) and the DFG (SFB1238 Control and Dynamics of Quantum Materials, 277146847, projects A04, B07 and C02).
    \end{acknowledgments}

\bibliography{Literature}
\clearpage
\pagebreak

\appendix
\renewcommand{\thefigure}{S\arabic{figure}}
\renewcommand{\thetable}{S\arabic{table}}
\setcounter{figure}{0}
\setcounter{table}{0}

\section{Supplementary Material}\label{sec:appendix}
\subsection{Breakdown in DC and fitting with Joule heating model}
\begin{figure}
        \centering
        \sidesubfloat[\centering]{{\includegraphics[height=4cm,keepaspectratio=true]{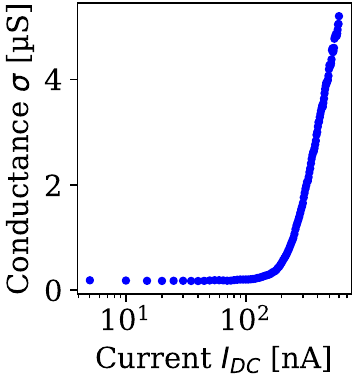} }}
            \sidesubfloat[\centering]{{\includegraphics[height=4cm,keepaspectratio=true]{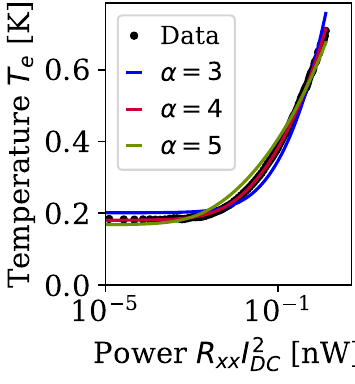} }}
        \caption{\justifying\textbf{Breakdown in DC and fitting of joule model} Left: Conductance $\sigma$ as a function of the applied current $I_{\rm DC}$, which is inferred from a locking measurement of the longitudinal resistance at $B=\SI{2}{\tesla}$ and $T\approx\SI{20}{\milli\kelvin}$. Right: Electronic temperature $T_e$ as a function of the applied power $R_{xx}I_{\rm DC}^2$ is shown in black dots. The fits with our joule heating model with varying exponent alpha are shown as solid lines, confirming that $\alpha=4$ shows a good agreement with data.}
        \label{fig:DC_breakdown}
    \end{figure}

    To analyze the breakdown under DC bias, we measured the longitudinal conductance $\sigma$ as a function of the applied current $I_{\rm DC}$ at base temperature. The data, shown in Fig.\ref{fig:DC_breakdown}a, displays a clear onset of bulk conduction above a current threshold $I_{\rm DC}\simeq\SI{20}{\nano\ampere}$. In Fig.\ref{fig:DC_breakdown}b, we extract the electronic temperature using the VRH calibration curve and plot it against the dissipated power $R_{xx} I_{\rm DC}^2$ as in Fig.\ref{fig:temp}b. We fit the Joule heating model with different values of the exponent $\alpha$. We find that the value $\alpha = 4$ gives the best agreement with the experiment. Plots for $\alpha=3$ and to \num{5} are also shown for comparison.

\subsection{\label{sec:corbino} Measurement of bulk current in Corbino geometry}
    To confirm that the increase of resistance is due to the onset of bulk conductivity, we fabricated a device in a Corbino geometry as shown in Fig.~\ref{fig:corbino-calib}a. The Corbino has an inner diameter of \SI{1}{\milli\meter} and an outer diameter of \SI{1.2}{\milli\meter}. The structure features 4 RF ports that are capacitively coupled to the edges of the V-BST film on the top left in Fig.~\ref{fig:corbino-calib}a. They are connected via coplanar waveguides to the RF setup. The remaining contacts are ohmic and are used to characterize the DC properties of the sample. The nano-fabrication is analogous to the hall-bar samples. We measure the current $I_{\rm bulk}$ that drains to the inner edge of the Corbino device, applying a current of \SI{1}{\nano\ampere} to the outer edge as sketched in Fig.~\ref{fig:corbino-calib}a. Although voltage bias is typically used in Corbino geometries, we note that a current bias is here achievable due to the presence of a finite residual bulk conductance. The resulting breakdown measurement is shown in Fig.~\ref{fig:corbino-calib}b), which shows the current $I_{\rm bulk}$ as a function of the microwave amplitude $U$ for frequencies from \SI{1}{\giga\hertz} to \SI{24}{\giga\hertz}, analog to Fig.~\ref{fig:breakdown}c). We see the same frequency and amplitude dependence as for the longitudinal resistance, confirming that the observed behavior is due to an onset of bulk conductivity.

\subsection{\label{sec:power_calib} Calibration of microwave amplitude}
\begin{figure*}[t]
            \centering
            \sidesubfloat[\centering]{{\includegraphics[height=4.2cm,keepaspectratio=true]{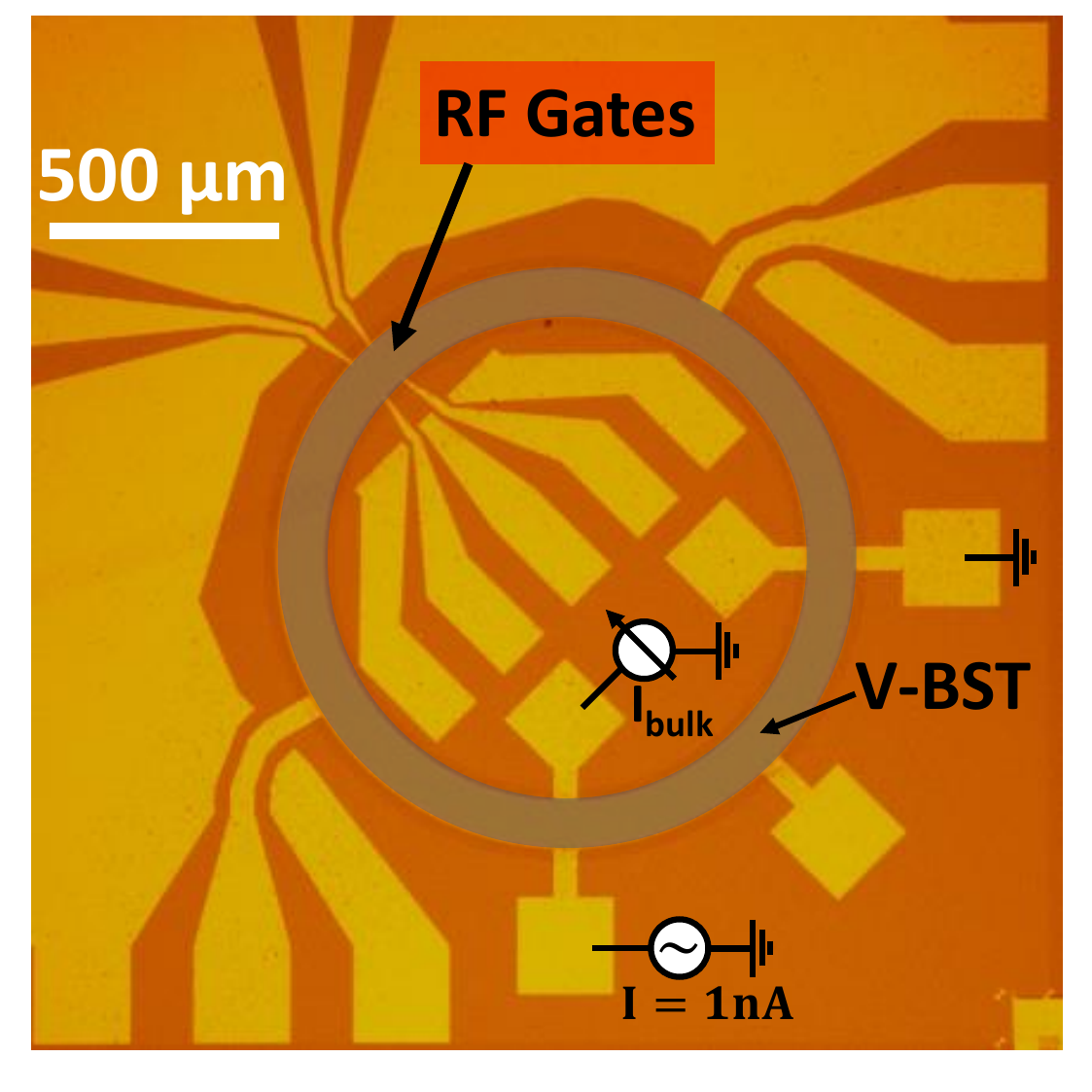} }}
            \sidesubfloat[\centering]{{\includegraphics[height=4.5cm,keepaspectratio=true]{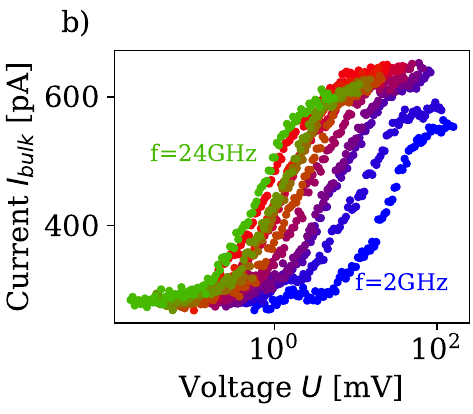}}}\quad
            \sidesubfloat[\centering]{{\includegraphics[height=4.5cm,keepaspectratio=true]{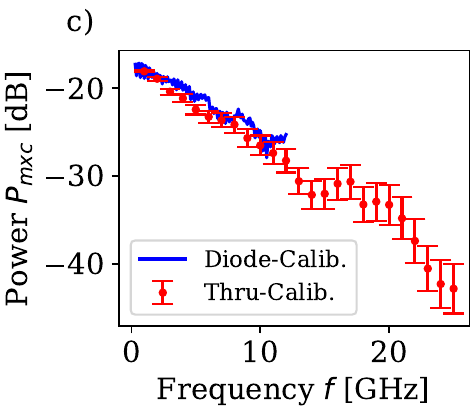} }}\quad
            \caption{\justifying\textbf{Corbino Geometry and Calibration:} a) Microscope image of Sample B in an RF Corbino geometry. The mesa is false-colored in dark yellow, the RF finger gates are placed at the top left and all other contacts are ohmic. The Corbino ring has a width of \SI{100}{\micro\meter}. b) Direct Measurement of the current that is drained to the inner ground in the Corbino geometry, as a function of the microwave amplitude $U_{\rm RF}$ for frequencies from \SI{1}{\giga\hertz} to \SI{24}{\giga\hertz} c) Calibration measurement of the RF power at the sample as function of the frequency, using a Through-Calibration (red) and a diode calibration (blue).}
            \label{fig:corbino-calib}
        \end{figure*}

    To calibrate the RF power $P_{\rm in}$, we measure an additional device, which connects the RF ports directly with an Au coplanar waveguide deposited on the InP substrate. We use a 50 \si{\ohm}-matched RF setup with identical input and output coaxial lines. This allows us to determine the power at the sample by measuring the transmission $S_{thru}$:
    \begin{equation}
        S_{thru}=S_{in} \cdot S_{out}
    \end{equation}
    where $S_{in/out}$ are the transmission coefficient of the input and output lines respectively. Assuming $S_{in}=S_{out}$, we apply the following correction factor for the microwave amplitude:
    \begin{equation}
        U_{\rm RF}=\sqrt{S_{thru}}\ U_{0} = S_{in}\ U_{0}
    \end{equation}
    with $U_{0}$ being the amplitude of the VNA output. 
    We make a coarse estimation of error by using the loss-coefficient of the cables which are given by approx. \SI{0.04}{\dB\per\meter\per\giga\hertz} and the cable length of \SI{4}{\meter}, which is shown as error bars in Fig.~\ref{fig:corbino-calib}c) and is also used for the error bars in the main manuscript.
    Since the calibration plays a key role in this work, we also performed an independent calibration measurement by using a tunnel diode detector (\textit{Herotek DT1-40}) to measure the power at the sample stage, which is shown in blue in Fig.~\ref{fig:corbino-calib}c). We confirmed by room-temperature measurements that the frequency-dependence of the sensitivity is minor and is therefore ignored.

\subsection{\label{sec:model_hyp} Model approximations and assumptions}

Our main model on heating of the electron temperature makes a number of strong approximations and assumptions, which we discuss more in detail in this section.\\

\begin{itemize}
\item {\sl \  Electron and phonon temperatures} -- The temperature parameter $T_e$ in the VRH equation (Eq. (2)) is strictly speaking both the electron temperature as well as that of the phonons which assist the hopping process. Using the equation here as a proxy to deduce the local electron temperature $T_e$ implicitly assumes that the phonons in the \SI{8}{\nano\meter}-thin film of V-BST have the same temperature $T_e$. In the bulk of the sample (\SI{350}{\micro\meter}-thick InP substrate), with much larger dimensions, we assume that the phonons remain in thermal equilibrium at $T_p$.

\item {\sl \ Homogeneity of the temperatures} -- We here assume for simplicity that the temperatures $T_e$ and $T_p$ are uniform over the whole sample. This neglects notably variations of $T_e$ in particular near the corners of the device where the contact resistances dissipate \cite{Ferguson2025}. Differences of the bottom and top surface of the topological insulator layer are not taken into account as their distance (\SI{8}{\nano\meter}) is much shorter than the phonon wavelength in the temperature regime of the experiment.

\item {\sl \ Geometry of the microwave field} -- The model does not account for the complex geometry of the microwave field, which is stronger near the excitation finger gate, and also varies with the excitation frequency $\omega$. We simply model the whole process with a single effective dissipation parameter $G(\omega)$.

\item {\sl \ Energy transport } -- The energy transport is described by a purely phenomenologically model only with two parameters, $\alpha$ and $\Sigma$. The model does, for example, not take into account that phonons at longer and shorter wavelength may be affected by different scattering mechanisms.

\end{itemize}

\subsection{Discussion of other mechanisms}

\begin{figure*}[t]
    \centering
    \includegraphics[width=0.8\textwidth]{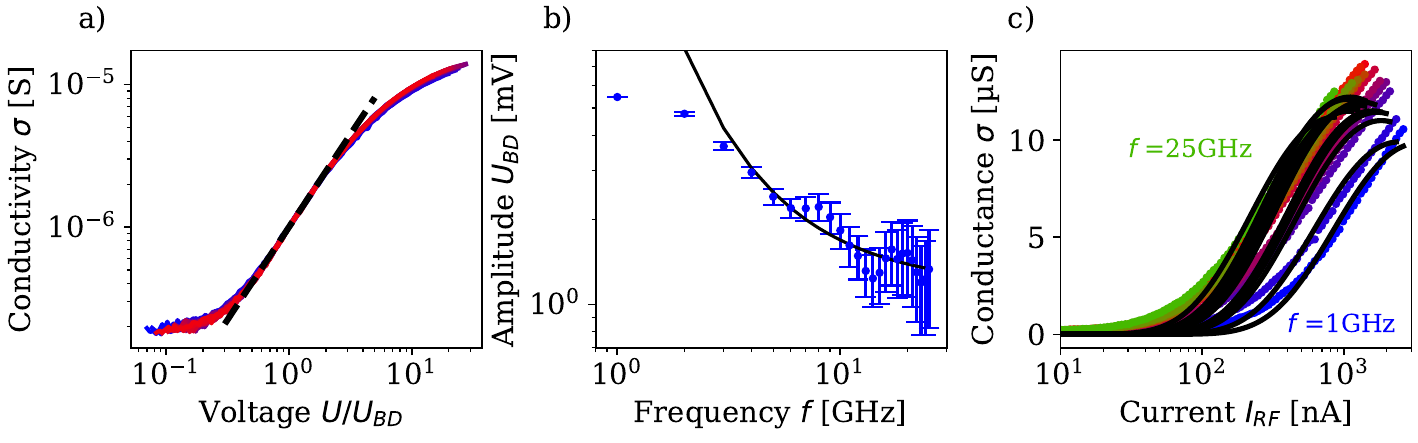}
    \caption{\justifying\textbf{Comparison with other mechanisms:} a) Amplitude dependence of the conductance, rescaled by the breakdown amplitude $U_{\text{BD}}$ for all frequencies from 1 to \SI{25}{\giga\hertz}. The dashed lines correspond to a power law of $U^{1.3}$. b) Extracted breakdown amplitudes $U_{\text{BD}}$ as a function of frequency $f$. The black line corresponds to a fitting with the expected frequency dependence of photon-assisted hopping, yielding $\omega_c=\SI{13(2)}{\giga\hertz}$. c) The conductance as a function of the applied current $I_{RF}$ (estimated as $I_{RF}\approx U/R_K$) for frequencies ranging from 1 to \SI{25}{\giga\hertz}. The black lines correspond to a Fit with the VRH law in Eq.~8.}
    \label{fig:other_mechanism}
\end{figure*}

As mentioned in the main manuscript, we also considered two alternative mechanisms that could, in principle, contribute to the observed breakdown. First, we explored photon-assisted hopping, where the absorbed microwave photons play a role analogous to phonons in classical variable-range hopping (VRH) by assisting the hopping between neighboring sites. Second, we considered a mechanism driven by the electric field, as discussed in the context of quantum Hall transitions by Polyakov and Shklovskii \cite{Polyakov1993}. In the following, we briefly outline the photon-assisted hopping model and explain why it does not agree with our data.

\subsubsection{Photon-assisted hopping conductivity}
    
    To describe the frequency and amplitude dependence of the conductance under microwave irradiation, we follow a model based on the Efros-Shklovskii framework for VRH \cite{efros_1975}, extended to include photon-assisted transitions, as proposed by Zvyagin \cite{zvyagin_hopping_1978}. We first note that transitions assisted only by photons play a marginal role due to the monochromatic nature of the AC field, which does not allow for matching energies, as pointed in Ref. \cite{zvyagin_hopping_1978}. We therefore consider the following processes where $n$ photons and 1 phonon are involved. For such processes, the hopping probability $\Gamma_{ij}$ between localized sites $i$ and $j$ in an AC electric field is given by:
    \begin{equation}
        \Gamma_{ij} \propto e^{\frac{-2|r_{ij}|}{\xi}}  (1 - f_j)  f_i  \sum_{n=-\infty}^{\infty} J_n^2\left(\frac{e E r_{ij}}{\hbar \omega}\right)  N(\Delta E_{ij} - n \hbar \omega)
    \end{equation}
    where $\xi$ is the localization length, $r_{ij}$ the hopping distance, and $f_i$, $f_j$ are Fermi-Dirac occupation probabilities. The Bessel function $J_n$ describes the probability of absorbing or emitting $n$ photons, and $N(\Delta E)$ is the Bose-Einstein factor that accounts for the phonon process needed to ensure energy conservation.
    
    In the absence of a driving field, we verified that the model recovers the usual Mott law of VRH in 2D:
    \begin{equation}
        \sigma(T) = \frac{\sigma_0}{T} \exp\left[-\left(\frac{T_M}{T}\right)^{1/3}\right].
    \end{equation}
    It is important to note that the Mott VRH law differs from the Efros-Shklovskii framework for VRH \cite{efros_1975}, which we used earlier to fit the temperature dependence of the conductance. However, at low temperatures, both models yield a similar functional form and describe the transport reasonably well.
    
    In the presence of an AC field, at low temperatures ($k_B T \ll e E r_{ij}, \hbar \omega$), hopping is dominated by transitions between sites with energy difference $\Delta E_{ij} \approx n\hbar \omega$. In this regime, the occupation factors simplify to $f_i(1 - f_j) \approx 1$. Furthermore, single-photon processes ($n = 1$) dominate in the low-field limit ($e E r_{ij} \ll \hbar \omega$). In this regime, we use the critical path method \cite{zvyagin_hopping_1978,Keiper1978} to find the most likely hopping distance $r_c$. Assuming a constant density of states $\rho_0$ (a Coulomb gap yielded similar results), the average distance between such resonant sites is:
    \begin{equation}
        \pi r_c^2 \rho_0 \hbar \omega = 1 \quad \Rightarrow \quad r_c = \left(\pi \rho_0 \hbar \omega\right)^{-1/2}.
    \end{equation}
    
    Based on the above-mentioned assumption, the model then predicts the following behavior:
    \begin{itemize}
        \item \ The conductance should scale as $\sigma \propto U^2$, i.e., linearly with absorbed microwave power.
        \item \ The breakdown amplitude $U_{\text{BD}}$ should depend on frequency as: $U_{\text{BD}} \propto \exp\left(0.5 \sqrt{\omega_c / \omega}\right)$, 
        where the cutoff frequency is defined as $\omega_c = 1 / (\pi \rho_0 \hbar \xi^2)$.
        \item \ From $\omega_c$, we can estimate the product $\rho_0 \xi^2$, and from this derive a corresponding Mott activation temperature: $T_m = \frac{1}{k_B \rho_0 \xi^2}$. 
    \end{itemize}
    
    In Fig.~\ref{fig:other_mechanism}a and ~\ref{fig:other_mechanism}b, we compare these predictions to our data. Panel a) shows the amplitude dependence of the conductance, where the voltage axis is rescaled by $U_{\text{BD}}$ to allow comparison across frequencies. The curves approximately collapse onto a single line, but the slope in the log-log plot is around \num{1.3}, which is lower than the expected value of 2. Additionally, deviations are observed at both low and high microwave powers, suggesting that the simple power-law model does not fully capture the behavior.
    
    The frequency dependence of $U_{\text{BD}}$, shown in Fig.~\ref{fig:other_mechanism}b, fits moderately well the predicted exponential relation only at higher frequencies. The extracted cutoff frequency is $\omega_c \approx 2\pi \cdot \SI{16(2)}{\giga\hertz}$, which fails to describe the saturation behavior at low frequencies seen in our measurements.
    
    Finally, using this value of $\omega_c$, we calculate the corresponding activation temperature: $
    T_m = \pi \cdot \frac{\hbar \omega_c}{k_B} \approx \SI{2}{\kelvin}$
    which is significantly lower than the experimentally extracted VRH activation temperatures, such as $T_M = \SI{286(7)}{\kelvin}$ (please note that the activation temperature obtained here with Mott's law is different from the one in the Efros-Shklovskii framework we employ in the main text \cite{efros_1975}). This discrepancy further indicates that the photon-assisted hopping model does not adequately describe the dominant mechanism in our system: we have been unable to find unique parameters that fit both the RF and DC response.

    \subsubsection{Electric field driven breakdown}
    To verify if an electric-field-driven mechanism could explain our data, we now consider the model by Polyakov and Shklovskii \cite{Polyakov1993}. In their work, they describe hopping transport in strong electric fields by introducing an effective electronic temperature $T_{\rm eff}$, which increases with the Hall electric field $E_H$, which is proportional to the applied current $I$. Hence, current plays a similar role as temperature in standard variable-range hopping and we can define the conductance as:
    \begin{equation}
        \sigma=\frac{\sigma_{0,I}}{I} e^{-\sqrt{I_0/I}}
        \label{eq:polyakov}
    \end{equation}
    Compared to Joule heating, which gives an effective temperature scaling of $T_{{\rm eff}} \propto I^{2/\alpha}$, the two models differ only by the exponent. However, since we consistently find an exponent of $\alpha \gtrsim \num{3.4}$, our observation is not well described by the model of Polyakov and Shklovskii. For completeness, we show fits of the conductance as a function of the amplitude and frequency of the microwave drive with Eq.~\ref{eq:polyakov} (see Fig.~\ref{fig:other_mechanism}c), exhibiting significant differences with our data. This confirms that this model does not describe the amplitude dependence well.

\subsection{Other samples and reproducibility}

To validate the generality and reproducibility of our findings, we performed the same measurements and analysis of the breakdown under microwave fields on additional samples: Sample B, a Corbino device, and Sample C, a Hall bar device. Both devices exhibit robust QAH quantization at low temperatures with $R_{yx} = h/e^2$, and show consistent breakdown signatures under microwave excitation.

\begin{figure*}[h]
    \centering
    \includegraphics[width=\textwidth]{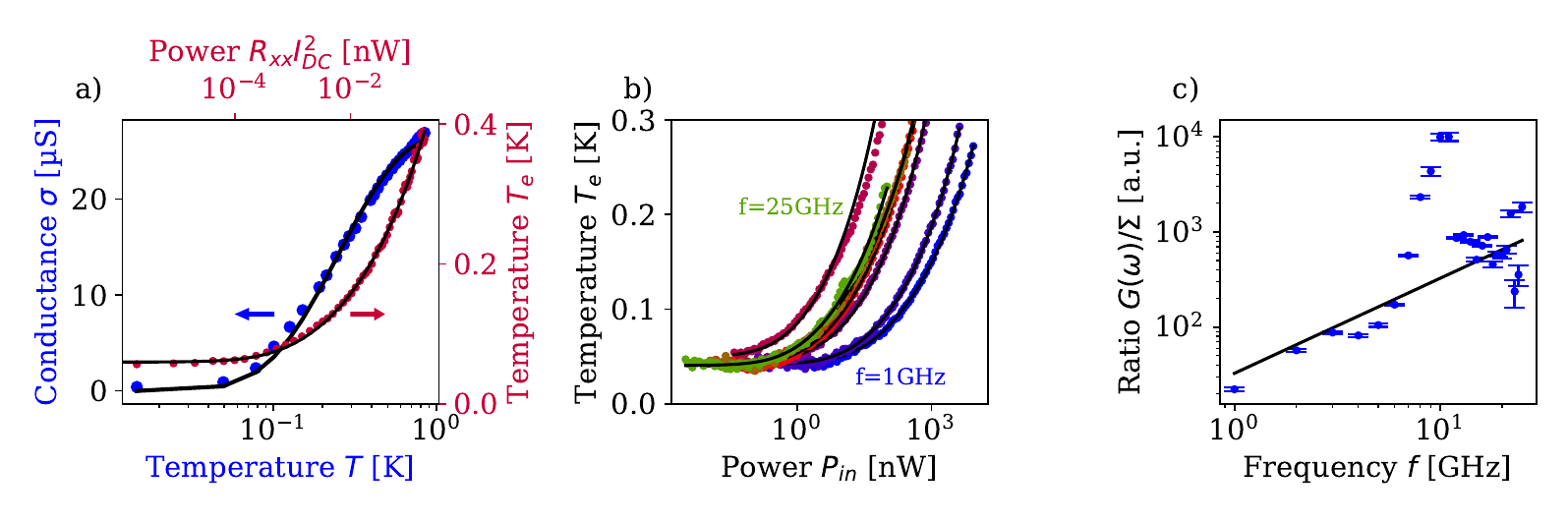}
    \caption{\justifying\textbf{Joule heating analysis of Sample B (Corbino geometry):} 
    a) Bulk conductance $\sigma$ as a function of phonon temperature $T_p$. The blue curve is fitted with the VRH model, yielding $T_0 = \SI{4.0(0.1)}{\kelvin}$ and used to calibrate the electron temperature $T_e$. The red curve shows $T_e$ extracted from DC breakdown and fitted with the Joule heating model, giving $\Sigma = \SI{3.9(0.1)}{\nano\watt\per\kelvin^\alpha}$ and $\alpha = \num{3.2\pm0.1}$. 
    b) $T_e$ as a function of microwave power $P_{\text{in}}$ at different frequencies. Fits to the Joule model yield $\alpha = \num{3.4\pm0.2}$. 
    c) Frequency dependence and fit of the parameter $G(\omega)/\Sigma=A \omega^s$, yielding $A=\SI{322(413)}{\kelvin^\alpha\siemens\per\giga\hertz^s \per\watt}$ and $s=\num{0.8\pm0.4}$}
    \label{fig:jul5a}
\end{figure*}

For Sample B, shown in Fig.~\ref{fig:jul5a}a--c, we use the same Joule heating model as for Sample A and extract a variable-range hopping (VRH) activation temperature of $T_0 = \SI{4.0(0.1)}{\kelvin}$. The electron temperature $T_e$ extracted from the DC breakdown (Fig.~\ref{fig:jul5a}a, red curve) fits well with the Joule heating model, yielding $\Sigma = \SI{3.9(0.1)}{\nano\watt\per\kelvin^\alpha}$ and $\alpha = \num{3.2\pm0.1}$. The microwave-induced breakdown (Fig.~\ref{fig:jul5a}b) is also well described by the Joule heating model, resulting in $\alpha = \num{3.4\pm0.2}$. The frequency dependence of the dissipation ratio $G(\omega)/\Sigma$ is again fitted by a power law, yielding a slightly lower exponent $s\approx\num{0.8}$. This confirms that the breakdown mechanism is reproducible in the Corbino geometry and bulk-dominated. 

\begin{figure*}[h]
    \centering
    \includegraphics[width=\textwidth]{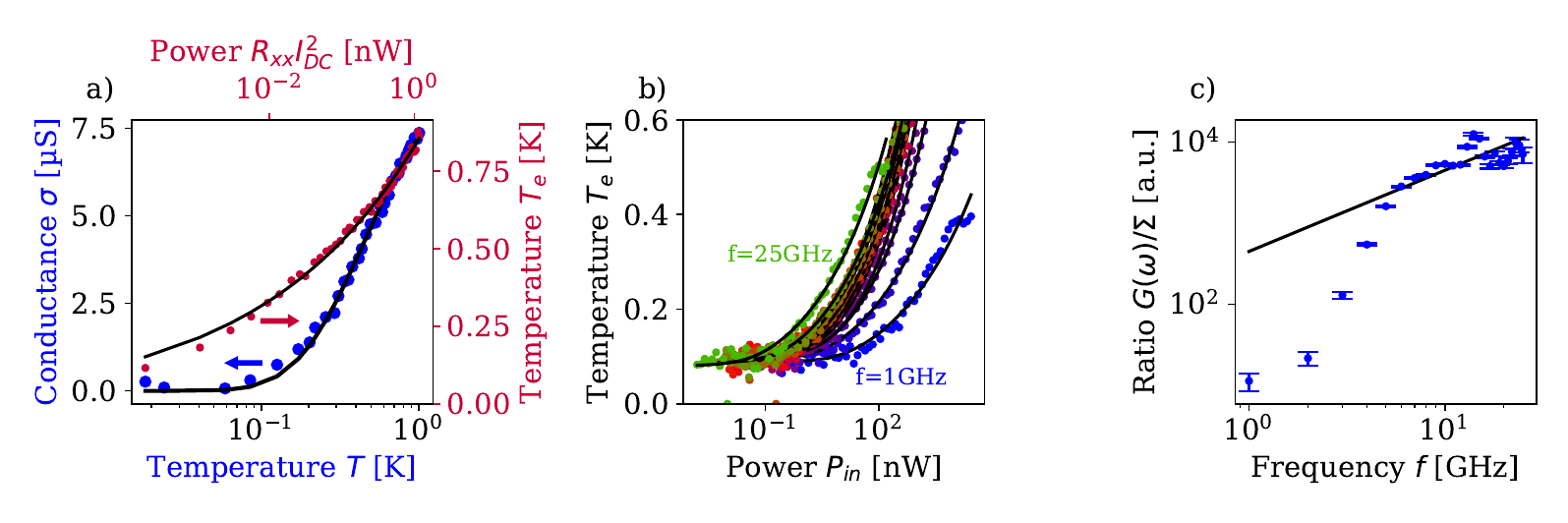}
    \caption{\justifying\textbf{Joule heating analysis of Sample C (Hall bar geometry):} 
    a) Bulk conductance $\sigma$ as a function of temperature $T_p$. The VRH fit to the blue curve ($U = 0$) yields $T_0 = \SI{7.3(0.1)}{\kelvin}$, used to extract $T_e$. The red curve shows $T_e$ under DC current, fitted with the Joule model, yielding $\Sigma = \SI{2.6(0.1)}{\nano\watt\per\kelvin^\alpha}$ and $\alpha = \num{5.0\pm0.1}$. 
    b) $T_e$ versus RF power $P_{\text{in}}$ at various frequencies, fitted with the heating model and yielding $\alpha = \num{4.0\pm0.5}$. 
    c) Frequency dependence and fit of the parameter $G(\omega)/\Sigma=A \omega^s$, yielding $A=\SI{812(540)}{\kelvin^\alpha\siemens\per\giga\hertz^s \per\watt}$ and $s=\num{0.8\pm0.2}$}
    \label{fig:jul31a}
\end{figure*}

The results for Sample C are shown in Fig.~\ref{fig:jul31a}. Again, the Joule heating model accurately reproduces the DC and RF breakdown data across a broad frequency and power range. From the temperature-dependent conductance (Fig.~\ref{fig:jul31a}a), we extract a VRH activation temperature of $T_0 = \SI{7.3(0.1)}{\kelvin}$. The DC breakdown (red curve) fits well with the heating model, yielding $\Sigma = \SI{2.6(0.1)}{\nano\watt\per\kelvin^\alpha}$ and $\alpha = \num{5.0\pm0.1}$. The RF-induced breakdown data (Fig.~\ref{fig:jul31a}b) is also consistent with the Joule heating model and gives $\alpha = \num{4.0\pm0.5}$. However, the ratio $G(\omega)/\Sigma$ (Fig.~\ref{fig:jul31a}c) shows a different frequency dependence, particularly at low frequencies where dissipation is strongly suppressed. This highlights the challenge of precise frequency calibration, though we consistently observe enhanced dissipation at higher frequencies.

These results confirm that the thermal breakdown model is robust across different device geometries (Hall bar and Corbino), disorder strengths, and cooldowns. However, we note that the exponent $\alpha$ appears to be sample-dependent and does not show universal behavior.

\end{document}